\input{aipcheck}

\documentclass[
    ,final            
  ]
  {aipproc}

\layoutstyle{8x11single}

\begin{document}

\title{Do flares in Sagittarius A* reflect the last stage of tidal capture?}


\author{A. \v Cade\v z}{
  address={Faculty of Mathematics and Physics, University of Ljubljana, Jadranska 19, 1000 Ljubljana, Slovenia}
}

\author{M. Calvani}{
  address={INAF, Astronomical Observatory of Padova, Vicolo dell'Osservatorio 5, 35122 Padova, Italy}
}

\author{A. Gomboc}{
  address={Faculty of Mathematics and Physics, University of Ljubljana, Jadranska 19, 1000 Ljubljana, Slovenia}
}

\author{U. Kosti\' c}{
  address={Faculty of Mathematics and Physics, University of Ljubljana, Jadranska 19, 1000 Ljubljana, Slovenia}
}

\begin{abstract}
In recent years the case for the presence of $3-4\times 10^6 M_\odot$ black hole in our Galactic Center has gain strength from results of stellar dynamics observations and from the detection of several rapid X-ray and IR flares observed in the Sagittarius A* from 2000 to 2004. Here we explore the idea that such flares are produced when the central black hole tidally captures and disrupts a small body - e.g. a comet or an asteroid.
\end{abstract}

\maketitle


\section{Sagittarius A* and observed flares}

Studies in the last decades have revealed presence of massive black holes in centers of most galaxies, 
including the quiet galactic nuclei, like the one in our Galaxy. The $3-4 \times 10^6 M_\odot$ black hole in SgrA* 
lies practically "in our backyard" and due
to its proximity is especially important for studying phenomena in the vicinity of massive black holes, 
including the effects of black hole's gravity on stellar systems. 

Recent observations of stars orbiting within 
1 pc from the Galactic center allowed for determination of the black hole's mass and provided evidence for the compactness 
of the central mass: S0-2 star has a periastron distance of 17 light hours \cite{Schodel} while S0-16 skimmed the 
SgrA* at only 45 AU or 6.2 light hours \cite{Ghez}, corresponding to $\sim 3000\, r_g$ and $\sim 1200\, r_g$, respectively
(where $r_g$ is the gravitational radius of the black hole: $r_g=GM_{bh}/c^2$). 
At such distances the stars are still safely outside the Roche radius: 
\begin{equation}
\rm{R_\mathcal{R}=50 \times (\varrho_{Solar}/\varrho_*)^{1/3} (10^6 M_{Solar}/M_{bh})^{2/3} r_g},
\end{equation}
which for a Solar type star in the Galactic center yields:  
$\rm{R_\mathcal{R} \sim 20\, r_g  \sim 7}$ light minutes, and therefore do not get tidally disrupted by the black hole \cite{Gomboc2005}, \cite{Gomboc2001}.
Nevertheless, it is interesting to note, that eccentricities of stellar orbits in the vicinity of Galactic center are all, except one, close to 1 (see Table 3 in \cite{Ghez}).

\begin{figure}
  \includegraphics[height=.7\textheight]{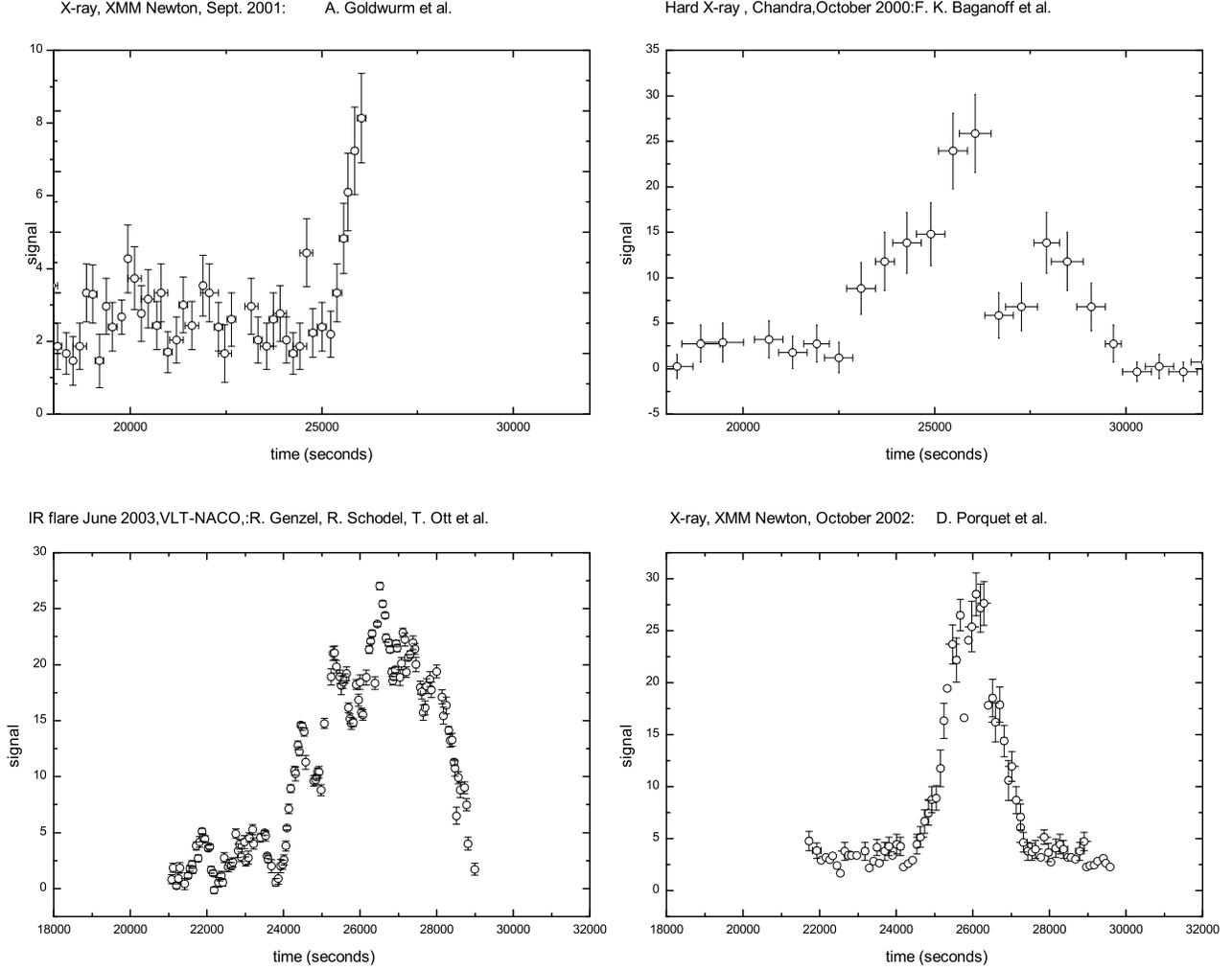}
  \caption{Light curves of X-ray and IR flares  observed coming from the Galactic Center in the period from October 2000 to June 2003. The signal has been rescaled with respect to the original data in order to better show "the time scale puzzle": the rise times of the flares are very similar; an exponential rise time of 900 s fits all the light curves quite well.}
\label{figFlares}  
\end{figure}

Additional evidence for the presence of a massive black hole in Galactic Center is coming from observed X-ray and IR flares in SgrA*. 
The first rapid X-ray flaring from the direction of Sgr $\rm{A^*}$ was observed by Chandra X-ray Observatory in October 2000, with the duration of about 10 ks \cite{Baganoff}. In September 2001, an early phase of a similar X-ray flare was observed by XMM-Newton. During this flare the luminosity increased by $\approx$ 20 in about 900 s \cite{Goldwurm}. The brightest (observed to date) X-ray flare, which reached a factor of 160 of the Sgr $\rm{A^*}$ quiescent value, was detected in October 2002 by XMM-Newton, and had a duration of 2.7 ks \cite{Porquet}. In May and June 2003 four infrared flares from Sgr $\rm{A^*}$ were observed with VLT. They had the duration from $\leq$ 0.9 ks to 5.1 ks and reached a variability factor of $\sim~$ 5 \cite{Genzel}. In addition, simultaneous observations with VLT and Chandra X-ray Observatory in July 2004 detected 5 events at infrared wavelengths and one moderately bright flare in X-ray domain, lasting about 2.5 ks with the brightening by about factor of $\sim$ 10 \cite{Eckart}. Light curves of some of the  observed flares are reproduced in Fig.~\ref{figFlares}.

\section{The time scale puzzle}
Figure \ref{figFlares} presents the following puzzling observations:
\begin{itemize}
\item
The rise and switch off rates of all flares are very similar.
\item
The rise-switch-off time scale is about 900 s.
\item
The light curve is similar in all wavebands. 
\end{itemize}

\subsection{Radiation diffusion time of a light source ($\tau_{\rm diff}$)}

Consider the fastest rate at which a massive light source can be extinguished.
Let the source be a homogeneous sphere with density $\rm{\varrho}$, mass $\rm{M}$ and radius $\rm{R}$ made of the of most transparent material available -- hydrogen at a high enough temperature (so that its opacity is due only to Thomson scattering, $\rm{\kappa = 0.4\, cm^2/g}$), and assume that photons are loosing no energy when diffusively scattering to the surface. Then, the radiation diffusion time of the light source is:  

\begin{equation}
\tau_{\rm diff} = {3 \kappa\varrho R^2\over{c\pi^2}} = 2.7\times
10^5 \left(
 {m\over
 {M_
 {\rm
Moon } }} \right)^{2/3}
\left({\varrho\over{g/cm^3}}\right)^{1/3}\, {\rm s}.
\label{cooltime}
\end{equation}

Thus, in order for a body to cool in 900 seconds, it should be less massive than the moon or implausibly rarified. 

To explain the same timescale of these events, they should have been caused by objects of almost exactly the same mass. We find this explanation highly unlikely and suggest that the timing is not so much due to sources themselves, but to the space-time of the central Galactic black hole they are moving in.

\section{Infalling point particle: observed intensity in the orbital plane as a function of time and longitude of observer}
\begin{figure}
  \includegraphics[height=.3\textheight]{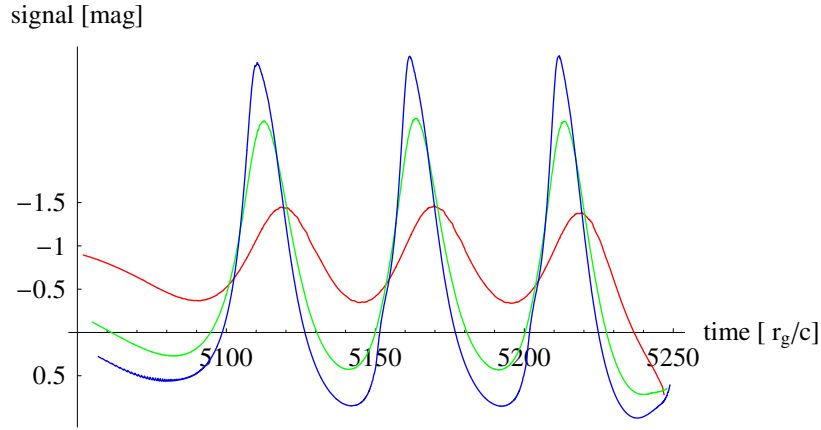}
  \caption{Light curves for infalling constant luminosity point particles; inclinations of the orbit with respect to observers are $16^\circ $ (red), $44^\circ$ (green) and $66^\circ$ (blue).}
\label{inclin}  
\end{figure}
\begin{figure}
  \includegraphics[height=.3\textheight]{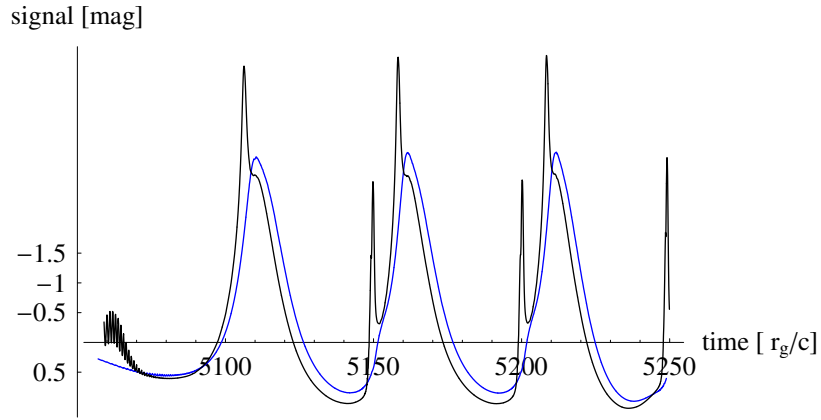}
  \caption{At high inclinations the primary and secondary image form Einstein rings and they are reflected in the light curve as short spikes; the two light curves are for observers at inclinations $66^\circ $ (blue) and $85^\circ$ (black).}
  \label{EinRIng}
\end{figure}

In order to understand the kind of phenomena that may be observable from the black hole "illuminated" by infalling material, we calculated theoretical light curves expected from very small, constant luminosity bodies falling to the black hole on a critical $l_{crit}=4 m {G M_{bh}\over c}$ orbit \cite{Kost}. Their apparent luminosity with respect to a far observer changes due to relativistic Doppler shift,  aberration of light and gravitational redshift \cite{And}. The reason for choosing critical orbits is the argument that most captures occur naturally from such orbits, since we expect that most captures are a result of a slow tidal process whereby orbital angular momentum of highly eccentric bound orbits is transferred to spin angular momentum until the orbital angular momentum approaches the critical value. The following three figures show samples of light curves expected from such events. 

\begin{figure}
  \includegraphics[height=.3\textheight]{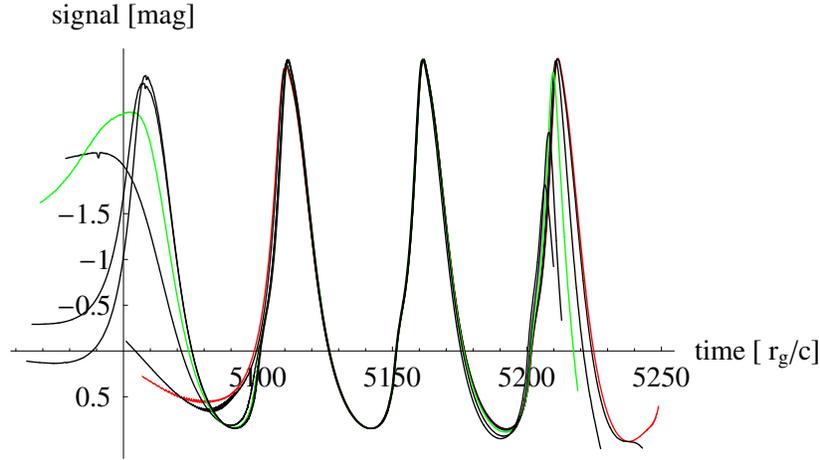}
  \caption{The observed light curve depends also on $\Omega$, the angle between the line of sight and the line of nodes. Six light curves belonging to $\iota=66^\circ $ orbits with $\Omega = ~19 ^\circ,~  47^\circ,~ 144^\circ,~206^\circ,~ 277^\circ$ and $ 310^\circ$ are shown here.}
  \label{Omega}
\end{figure}

Figure \ref{inclin} shows the dependence of the light curve on inclination of the orbit with respect to the observer. As expected, low inclinations produce less modulation. Extreme modulations can be seen only from high inclination orbits as shown in Fig.~\ref{EinRIng}. The two high peaks seen in the $\iota = 85^\circ$ light curve come from Einstein rings formed in the primary and secondary image. Figure \ref{Omega} emphasizes that the apparent light curve is sensitive also to the orientation of the line of nodes with respect to the line of sight ($\Omega$).  Here this dependence is illustrated for six $\iota = 66^\circ$ orbits. 

\section{Scenario}

\subsection{Galactic center}
\begin{figure}
  \includegraphics[height=.15\textheight]{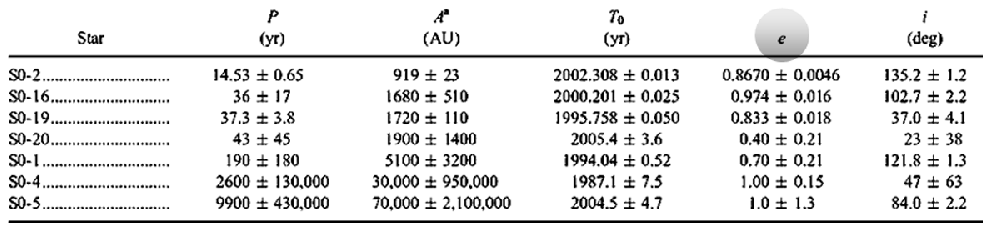}
  \caption{Orbital parameters for 7 stars moving around the Sgr $A^*$ black hole as determined by Ghez et al \cite{Ghez}. Note the high eccentricity of all the determined orbits.}
  \label{GC}
\end{figure}

Stars moving close to the Galactic center black hole are  tidally interacting with the black hole. This interaction promotes transfer of energy and angular momentum from orbital to internal. Since tidal forces most effectively do work on stellar or stellar-system modes resonant with respect to the characteristic time of periastron passage, the energy - angular momentum transfer is most effective for high eccentricity orbits having sufficiently short periastron crossing time. The result is slow decrease of orbital angular momentum and gradual increase of the eccentricity of stellar orbit. Since each next periastron is nearer than the previous, the tidal energy - angular momentum transfer accelerates and the stellar orbit eccentricity keeps increasing. The effect of tidal work on internal modes is their increase to the point of becoming unstable. This means that with decreasing periastron passage time nearer and nearer stellar satellites (planets, moons, asteroids, ...) are shaken off the star and become members of the larger black hole system. This view is based on a study of tidal interaction with a large central mass \cite{Gomboc2005}, and is supported by measurements of stellar orbits  parameters in the galactic center by Ghez et al \cite{Ghez} (see Fig.~\ref{GC}).

\subsection{Asteroids in the galactic center}
\begin{figure}
  \includegraphics[height=.4\textheight]{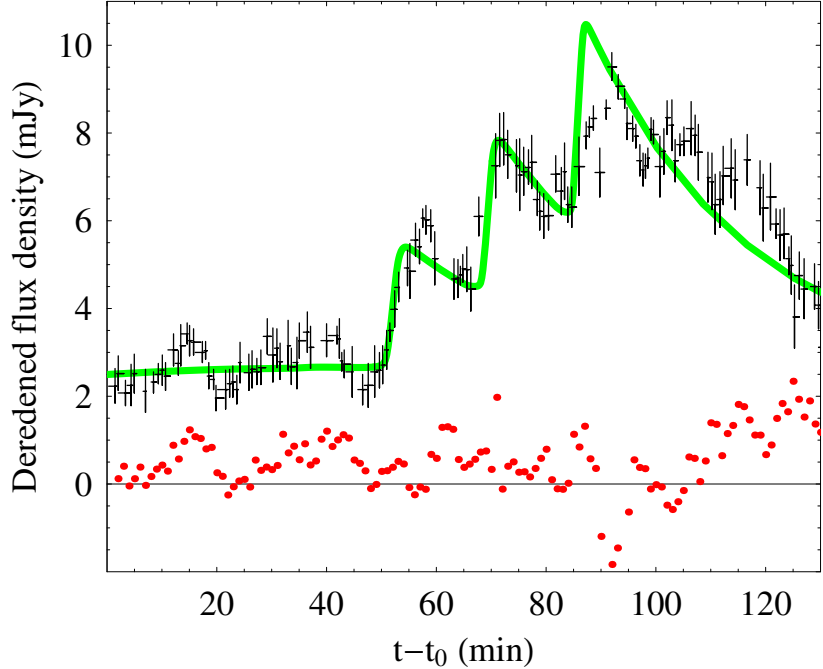}
  \caption{IR flare observed in Sgr $\rm{A^*}$ on June 16th, 2003 and our fit (green line) to the observed light curve obtained with a very rudimentary model, assuming that asteroid's luminosity is increasing exponentially with time and the luminosity of its tidal tail is decreasing exponentially with the distance from the asteroid's core. Red points are residuals after the fit.}
  \label{fit}
\end{figure}

We assume that it is not implausible to expect that galactic center stars were formed with a certain proportion of leftover material that was used in planets, moons, asteroids and comets. If the distribution of this material is similar to the distribution around the Sun, then, according to the above argument, the galactic center must be populated by a considerable number of solid objects such as planets, moons, asteroids and comets that move about the galactic center like comets in our own Kuiper belt. Many of the asteroids in this Kuiper belt move on highly eccentric orbits which reach deep into the potential 
well of the black hole. At their periastra they experience very high tides, which heat them and transfer angular momentum from orbit to spin. In this way the orbit is becoming more and 
more eccentric (parabolic) and the angular momentum is slowly approaching  the angular momentum of 
tidal capture $l=4 m {G M_{bh}\over c}$. The last tidal kick, that occurs just before capture, may release up to ~10\% mc$^2$ 
of tidal energy to the asteroid. This is more than enough to heat it to X-ray temperatures. Our model for the light curve of such an infalling asteroid is very simple. We assume that at a certain distance, only a few gravitational radia from the black hole, tides overwhelm the tensile strength of a solid asteroid and do enough work to evaporate it and heat it to X-ray temperatures. After evaporation, the asteroid is also tidally distributed along the orbit into a cometary like tail. In order to make the model as simple as possible, we assume that tidal heating is exponential with a characteristic heating time $\tau_h$ and we also assume exponentially decaying luminosity in the tidal tail with a characteristic scale $l_t$. The expected luminosity curve of such an event is then calculated by using the constant proper luminosity - point particle models described above as Green's functions. We take $\tau_h$ and $l_t$ as free parameters and adjust the mass of the black hole, the inclination of the orbit ($\iota$) and the line of nodes ($\Omega$) to obtain a synthetic light curve which best fits the June 2003 IR flash. The result of the fit (obtained from a limited data set of Green's function samples) is shown in Fig.~\ref{fit} and has been obtained for the following parameter values:
$$
\matrix{M_{bh} = 4 \times 10^6~M_\odot\cr
        \tau_h = 2300~ s \cr
        l_t=10^8~ km\cr
        \iota \approx 85^\circ \cr
        \Omega \approx 60^\circ}.
$$

\subsection{Time scales and energetics}

If one assumes that $\tau_h$ is (comparable to) the heating time of the evaporated but still dense asteroid ($\rho=2 g \ cm ^{-3}$), then eq.~\ref{cooltime}, would suggest the mass $m_{ast}\sim 4\times 10^{22}g$. The energy liberated by tidal work could be as high as: 
$\Delta E \sim 0.1 mc^2 \sim 4 \times 10^{42} erg$. We also note that our model needs the mass to be delivered to the vicinity of the black hole as a solid sufficiently dense piece; otherwise its Roche radius would be considerably higher and would make too long a tidal tail when reaching the critical orbit. 

A very rough estimate of capture rate could be: 
   stellar capture rate $\times$ number of asteroids per star: 
   $\sim (10^{-4} y^{-1}) × 10^5 = 10 y^{-1}$.

\section{Conclusions}                             
\begin{itemize}
\item{The light curve of tidal capture of an asteroid size body is modeled as the light curve of an almost point particle being captured and heated by tides on a critical orbit by the massive black hole. }
\item{The time scale fits the known mass of the central galactic black hole, }
\item{the expected energy release is compatible with observations, and }
\item{the observed frequency of events is consistent with the expected capture rate.}
\item{This simple model fits the observed light curve of the IR flare surprisingly well. }
\end{itemize}

\begin{theacknowledgments}
A\v C and AG acknowledge the support of Slovenian Research Agency. AG 
also acknowledges the receipt of Marie Curie European re-integration grant.
\end{theacknowledgments}

\end{document}